# Bias-Dependent Generation and Quenching of Defects in Pentacene

D. V. Lang[1,2], X. Chi[2], A. M. Sergent[3], and A. P. Ramirez[2,3]

[1]*Los Alamos National Laboratory, Los Alamos, NM 87545*
[2]*Columbia University, New York, NY*
[3]*Bell Laboratories, Lucent Technologies, 600 Mountain Avenue, Murray Hill, NJ 07974*

We describe a defect in pentacene single crystals that is created by bias stress and persists at room temperature for an hour in the dark but only seconds with 420nm illumination. The defect gives rise to a hole trap at $E_v + 0.38eV$ and causes metastable transport effects at room temperature. Creation and decay rates of the hole trap have a 0.67eV activation energy with a small ($10^8$ $s^{-1}$) prefactor, suggesting that atomic motion plays a key role in the generation and quenching process.

Defects and impurities play a dominant role in determining the properties of semiconductors. In the emerging area of organic semiconductors, however, little is known about them. One of the few known examples is tetracene as an impurity in anthracene, which forms a 0.42eV hole trap [1, 2]. Deep-level transient spectroscopy (DLTS) measurements of poly(p-phenylene-vinylene) (PPV) films [3] and polycrystalline pentacene films [4] indicate the presence of deep gap states. Calculations show that point defects in anthracene should give a trap level of about 0.28eV and dislocations should give a level of about 0.2eV [2, 5]. Northrup and Chabinyc [6] have recently calculated the gap state energies of various hydrogen- and oxygen-induced defects in crystalline pentacene and find hole traps at 0.34eV and 0.18eV from the valence band, respectively. They propose that these defects can be formed by bias stress and could potentially explain the metastable bias-stress effects in pentacene thin film transistors (TFTs) reported by Knipp et al. [7] Reynolds et al. [8] have also made passing reference to metastable effects, such as persistent photoconductivity, in pentacene thin films.

In this paper we will describe a defect in pentacene single crystals that can be generated by bias-stress and exhibits metastable effects in space-charge-limited current (SCLC) IV curves. The defect concentration recovers to zero-field equilibrium at room temperature in the dark in more than an hour, but in less than one second if illuminated with photons above the bandgap (2.2eV [9]). The defect gives rise to a hole trap at $E_v + 0.38eV$ that is generated with a 0.67eV activation energy when the Fermi level is closer than about 0.3eV to the valence band and disappears with the same activation energy when the Fermi level is above 0.4eV. We will discuss the experimental results and the various models that could explain the data.

The pentacene crystals were grown by physical vapor transport as described previously [10] and were 10-30mm$^2$ in area and 25-50μm thick. The e-beam evaporated contacts were Au (5000nm) on Ti (5nm) using a shadow mask. The array of 400μm-square metal pads had 30μm gaps along the x direction and 108μm gaps along y, allowing us to spatially map the characteristics of each crystal. The measurements were made on the 30μm gaps in a temperature-variable vacuum probe station with a Keithley 6517A electrometer. Photo-excitation was with a grating monochromator focused to an optical flux of 3 x 10$^{13}$ photons/cm$^2$sec at 420nm (2.95eV).

Figure 1 shows the SCLC IV curves under various conditions. The initial condition for the broken lines in Fig. 1 is the zero-field equilibrium state, which may be achieved in the dark at zero bias in several hours at room temperature, or more rapidly at 360K in one minute or with 2.95eV photons at room temperature in a few seconds. In Fig. 1 we used 2.95eV photons. All IV curves were from V=0 to a particular end-point voltage (30, 100, 300 or 600V) and consisted of 20 evenly spaced bias steps with a 5 sec dwell time. The initial scans define a typical SCLC power-law (I ~ V$^n$) [2, 11] with an initial exponent n=3.5 and an asymptotic exponent n=2 at high bias. Such a power law for n>2 is indicative of an exponential distribution of traps with a characteristic energy of (n-1)kT, where k is the Boltzman constant and T is the absolute temperature.

The initial condition for the solid lines in Fig. 1 is a bias-polarized state, which is achieved by holding the sample at the respective end-point voltage for 100sec. Each of the polarized scans was from V=0 to the respective end-point voltage after the polarization voltage was removed. Subsequent scans repeated within a few minutes of the polarized scan, but without additional polarization, behave the same as the first and are independent of the bias scan direction. All of the polarized IV curves in Fig. 1 have similar shapes – a polarization-dependent voltage threshold and a steep power law (n=11) increase in current that asymptotically joins the initial curve at the polarization bias. These bias-polarized curves display the classic shape of SCLC IV curves in material with traps [2, 11, 12] where the steady-state current changes rapidly as the Fermi level moves through a trap level as a function of bias voltage. The voltage threshold is a measure of the trap concentration, while the trap energy can be inferred from the activation energies of the current above and below the current step. For the 600V bias-polarized state, the activation energy is 0.55eV below the step and 0.21eV at 600V after the step. If we take the activation energy to be approximately equal to the Fermi level, the 0.38eV midpoint corresponds to the center of the band of traps. The trap concentration can be calculated [11, 12] from the voltage threshold (100V), trap energy (0.38eV), and power law (n=11) to be 2.5 x $10^{14}$ $cm^{-3}$ for a 0.1eV-wide distribution of trap energies. This is close to the maximum uniform space charge density (5 x $10^{14}$ $cm^{-3}$) that can be supported in such a structure at 600V [13]. Thus, the bias-polarized state corresponds to an increased concentration of defects with gap states centered at $E_v$ + 0.38eV, while the zero-field equilibrium state corresponds to the disappearance of these defects.

The trap dominated SCLC curves are only found, however, in the regions of our crystals where the low-field Fermi level is more than 0.4eV from the valence band so that the Fermi level can be swept through the trap level with increasing bias. Parts of our crystals have a Fermi level as low as 0.26eV where the IV step due to trap filling cannot be observed and there is no hysteresis in the IV curves, as shown in Fig. 1. The spatial distribution of Fermi levels results from unintended variations in p-type dopants across our crystals. The metastable curves in Fig. 1 also exhibit non-hysteretic conductivity when illuminated at 420nm; this photoconductivity also has an activation energy of 0.26eV.

We can study the dynamics of the transition between the bias-polarized state and the low-field equilibrium state by measuring current versus time at a fixed bias following different initial conditions. To optimize the dynamic range and signal-to-noise ratio, we have studied the initial build-up in the 600V bias-polarized state and the decay of this state at 140V. The decay transients for three temperatures (320, 340 and 360K) in Fig. 2a are measured at 140V immediately after 1000s at 600V and correspond to the transition from the 600V-polarized state to the 140V equilibrium state. The crosses on the decay curves in Fig. 2a denote the shift with temperature for the same relative point on each curve and are shown in the Arrhenius plot in Fig. 3. The polarization transients for three temperatures (300, 320 and 340K) in Fig. 2b are measured at 600V ($2x10^5$ V/cm) immediately after the end of the decay transient at 140V and correspond to the transition from the 140V equilibrium state to the 600V bias-polarized state. The open circles on these curves denote the inflection point that changes with temperature and are shown in Fig. 3. The current versus time during bias polarization has the same shape for other polarization voltages; thus the transition rate to the bias-polarized state does not depend on the magnitude of the electric field.

The temperature dependence of the transitions between the two limiting states is shown in Fig. 3. We have plotted the inverse of the times marked in Figs. 2a and 2b as the transition rate in the figure. These data are fit a function of the form r = Aexp(-E/kT), were r is the transition rate, A is the exponential prefactor, and E is the activation energy. The decay rate in Fig. 3 has E = 0.67eV and A = $9.0x10^7$ s$^{-1}$ while the polarization rate has E = 0.67eV and A = $1.05x10^9$ s$^{-1}$. The transitions in Fig. 2a&b are broader than a single exponential time constant, however, and correspond to a distribution of rates. The energies and prefactors in Fig. 3 roughly characterize the center of this distribution. Note that the activation energies for the decay and polarization rates are the same. If the metastable effects were due to the trapping and thermal release of holes at defect states in the gap, these energies would not be the same. Equal barriers for both directions of a reaction imply that the equilibrium energies of the final states are the same. This is true for two-level systems or atomic motion, but we are not aware of examples related to carrier trapping. Note also that the exponential prefactors are unusually small for thermal capture and emission of carriers at traps; the factor for the decay rate is $10^4$ smaller than the more typical trap emission prefactor of $10^{12}$ s$^{-1}$ [14].

Our results suggest a model where defects are created when the Fermi level is less than 0.3eV from the valence band and are quenched either thermally or optically when the Fermi level is more than 0.4eV from the valence band. The same 0.67eV activation energy for the defect generation and quenching process, along with the low ($10^8$ s$^{-1}$) prefactor, is consistent with a model where the rate-limiting step is the diffusion of weakly bound atoms. Before we discuss this model in more detail, however, we must rule out a more prosaic cause for the metastability we observe. Namely, a metastable electrostatic barrier in the sample might be created by defects that are positively charged by hole capture at the polarization bias. In this case Fig. 2b would correspond to hole capture and Fig. 2a to hole emission. Such a defect would need equal 0.67eV barriers for the capture and emission of holes (with a very low exponential prefactor), and also have no barrier to electron capture. We know of no examples of such a defect. We could imagine equal Coulomb barriers to hole capture and emission for a cluster of traps, such as at a grain boundary [15]. However, the lack of electric field dependence of our observed polarization rate rules out such a model. We could also consider a large lattice relaxation model, such as that which successfully describes persistent photoconductivity in III-V compound semiconductors [16]. This is unlikely, however, considering the weak Van der Waals bonding of the molecular crystal and the fact that intramolecular distortions are unlikely to give 0.67eV energy barriers [6]. Therefore, we believe that models involving carrier trapping and a metastable electrostatic barrier cannot describe our data.

On the other hand, we believe the pentacene defect reaction scenario proposed by Northrup and Chabinyc [6] gives important clues about the origin of our effect. This model is based on density functional calculations of various hydrogen- and oxygen-induced defects in crystalline pentacene. Adding H or OH to a pentacene molecule ($C_{22}H_{14}$) forms a four-fold-coordinated C atom that gives rise to three charge states in the gap (+/0/-). The capture and emission transitions among these states correspond to a deep donor level (+/0) at $E_v$ + 0.34eV and a deep acceptor level (0/-) at $E_v$ + 0.8eV. These defect states are due to the removal of the C atom from the ∂ bonding system, not the nature of the added atom. Therefore the gap states are essentially the same for either the addition of H (the C-$H_2$ defect) or OH (the C-HOH defect). For simplicity we will discuss the C-$H_2$ defect (a $C_{22}H_{15}$ molecule), although it should be kept in mind that the C-HOH defect ($C_{22}H_{15}O$) behaves essentially the same. A pentacene molecule with two C-$H_2$ defects ($C_{22}H_{16}$) is more stable than $C_{22}H_{15}$ and has no states in the gap. Therefore, the reaction

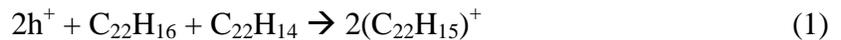

$$2h^+ + C_{22}H_{16} + C_{22}H_{14} \rightarrow 2(C_{22}H_{15})^+ \qquad (1)$$

is driven to the right as the Fermi level drops below $E_v$ + 0.34eV where the pairing energy is given by $1.04 - 2(0.34 - E_F)$ eV, with the Fermi energy $E_F$ measured from the valence band [6]. For 600V in our experiments $E_F$ = 0.21eV so the pairing energy reduces to 0.78eV.

Our observed range of defect energies ($E_v$ + 0.38 ± 0.05eV) is extremely close to the calculated energy ($E_v$ + 0.34eV) of the C-$H_2$ defect. In addition, the Fermi level dependence of our defect creation effect is remarkably close to the calculated energy of the positive charge state of the C-$H_2$ defect. Namely, we see no polarization hysteresis unless the Fermi level can be reduced below 0.3eV by the applied bias to create defects and returned above 0.4eV at low bias to observe the decay of the defect. However, the simple C-$H_2$ defect reaction in Eqn. 1 cannot adequately explain our 0.67eV activation barriers and the low exponential prefactors. Northrup and Chabinyc envision a model in Eqn. 1 where the C-$H_2$ defect is created by a proton jumping from a dihydropentacene ($C_{22}H_{16}$) molecule to a neighboring pentacene ($C_{22}H_{14}$). This would not give the symmetric generation and quenching activation energies that we observe because the equilibrium energies of the two sides of Eqn. 1 are very different. However, if the rate-limiting step in the process were the thermally activated diffusion of H, we would see symmetric barriers. In fact, activation energies below 1eV are typical of interstitial diffusion of impurities in inorganic semiconductors [17]. If the mobile H moves some distance before finding a suitable reaction site, the prefactor is of order $10^{12}$ s$^{-1}$ for each hop divided by the average number of hops. Since the observed prefactors are in the range $10^8 - 10^9$ s$^{-1}$, the average distance traveled is the square root of the number of hops ($10^3$ to $10^4$) times the distance of a single hop (assume ~5Å), which is approximately 30nm and corresponds to a reaction site concentration of order 5 x $10^{16}$ cm$^{-3}$. Finally, we need to explain the rapid quenching of our defects by above-bandgap light. The quenching conditions in our experiments correspond to C-$H_2$ defects being in the neutral charge state before illumination. The capture of a photo-generated electron into the $E_v$ + 0.8eV level of C-$H_2$ from the conduction band must dissipate 1.4eV, which is larger than the diffusion barrier of 0.67eV. If the electronic capture energy were channeled into the appropriate reaction coordinate by the well-known mechanism of recombination-enhanced diffusion [18, 19], this could lead to the greatly enhanced reaction rate we observe. We must stress, however, that this is merely a conjecture and the actual mechanism for the rapid photo-quenching of the defects is presently unknown.

In summary, we have described a bias-dependent defect generation and quenching effect in pentacene single crystals that gives rise to metastable transport effects at room temperature. Space charge limited current measurements in the bias-polarized state indicate an increase in the concentration of hole traps centered at $E_v + 0.38eV$ that persist for nearly an hour in the dark but can be rapidly removed by light above the 2.2eV bandgap. The creation and decay rates of this trap have the same activation energy (0.67eV) with a small exponential prefactor that is typical of atomic diffusion. Some aspects of our results can be explained by the dissociation of $C-H_2$ defect pairs, as proposed by Northrup and Chabinyc, but a detailed explanation of our results must take into account the apparent atomic diffusion that we observe. Our results suggest that four-fold coordinated C atoms with associated intermolecular H diffusion may be an important class of defect reactions in pentacene and, perhaps, other organic solids as well.

We wish to acknowledge stimulating discussions with M. S. Hybertsen and M. L. Steigerwald. We also acknowledge support from Los Alamos LDRD program as well as the Department of Energy, Basic Energy Sciences, grant number FWP# 04SCPE389.

**Figure Captions**

Figure 1.  Space charge limited current (SCLC) versus bias voltage showing metastable bias-polarization effects for four voltages (30, 100, 300 and 600V).  The additional two curves are 420nm photoconductivity for the same contact location and SCLC for a different location with a 0.26eV activation energy.  The latter two curves show no hysteresis.

Figure 2.  Current versus time at fixed bias for different initial conditions and temperatures.  (a) Decay of the 600V bias-polarized state at 140V for three temperatures. The cross marks on the curves are at the same fraction of the respective maxima and form the basis for the decays points in Fig. 3.  (b) Buildup of the 600V bias-polarized state at 600V after equilibrium at 140V for three temperatures.  The open circles on the curves are at the intersection of power-law fits for short and long times and form the basis for the polarization points in Fig. 3.

Figure 3.  Transition rate versus inverse temperature for the decay and polarization data in Fig. 2.  The fits to these data indicate the activation energy and exponential prefactor for each process.

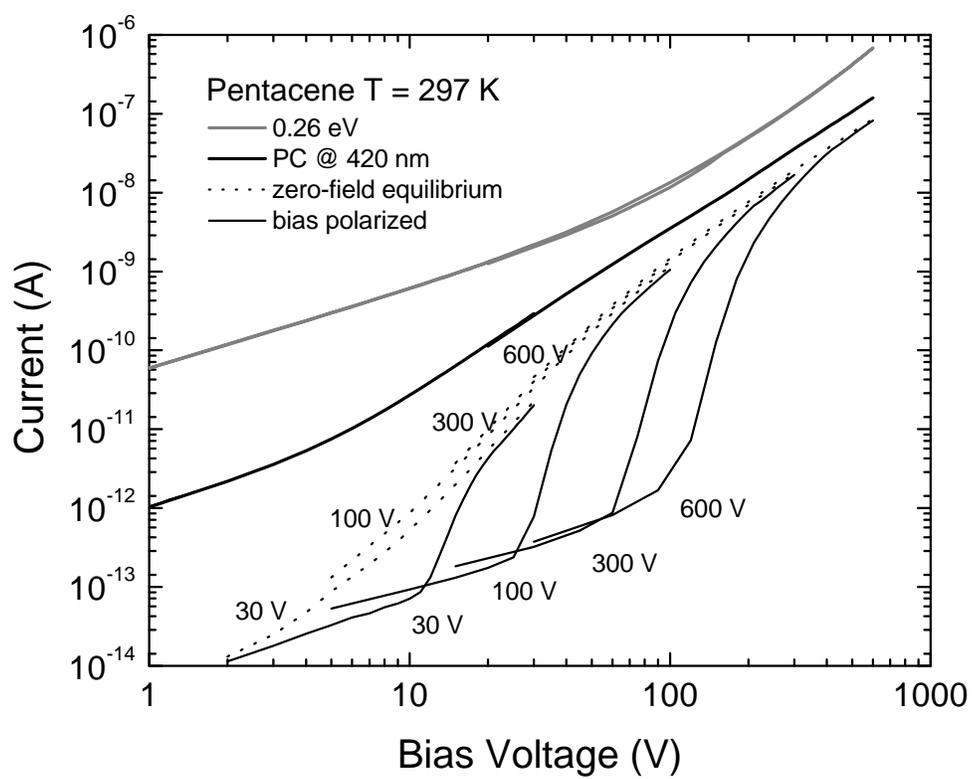

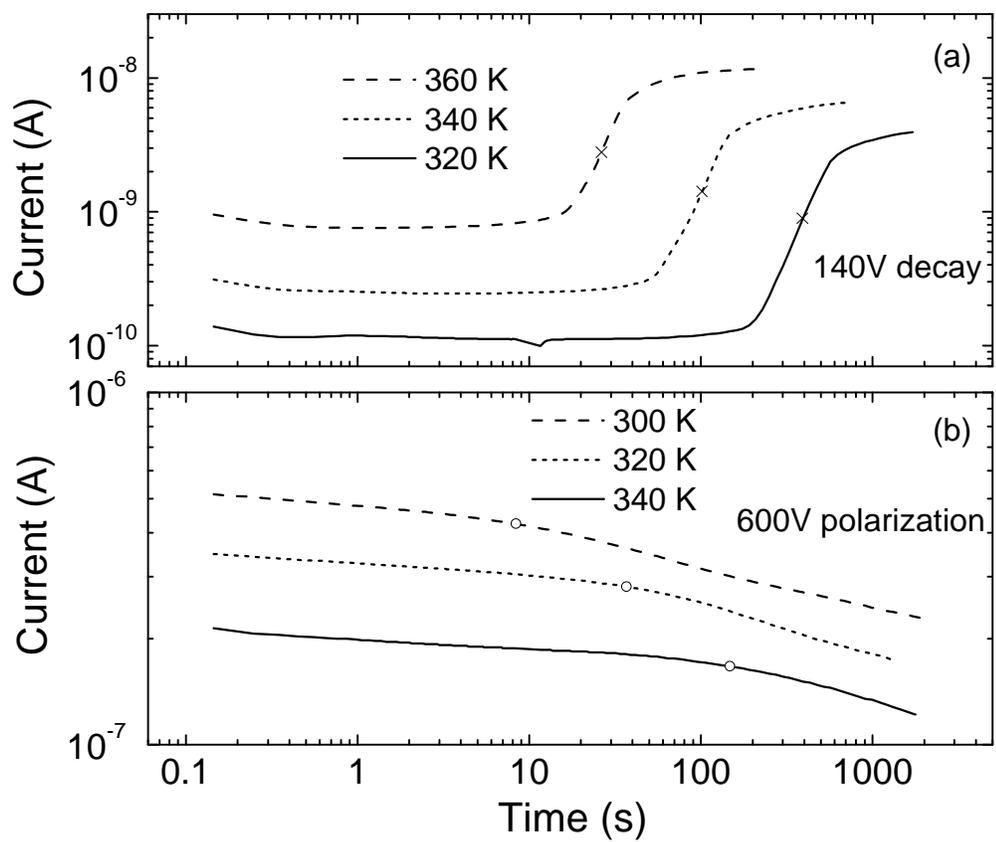

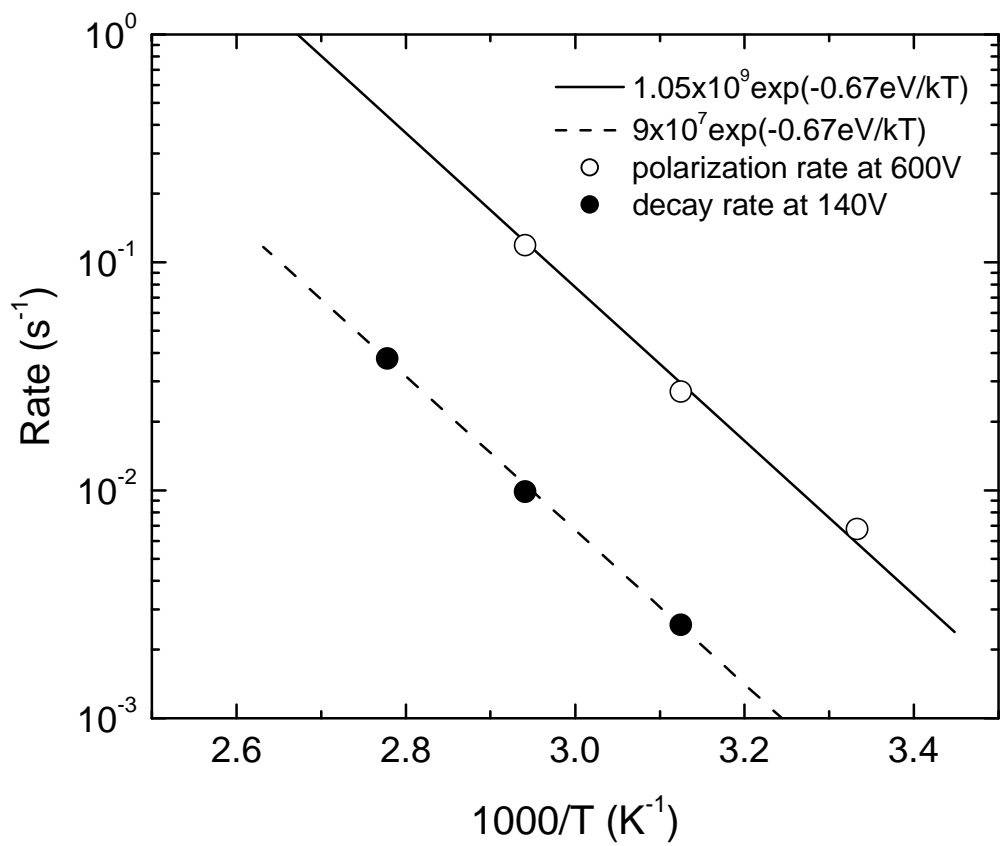